\documentclass[a4paper,11pt]{article}
\pdfoutput=1 

\usepackage[utf8]{inputenc} 
\usepackage{bbold}
\usepackage{slashed}
\usepackage{a4wide}
\usepackage{amsmath}
\usepackage{cite}       
\usepackage{xcolor}
\usepackage{amssymb}
\usepackage{amsfonts}
\usepackage{hyperref}
\usepackage{booktabs}
\usepackage{makecell}
\usepackage{multirow}
\usepackage{pdflscape}
\usepackage{textcomp}
\usepackage{gensymb}
\usepackage{bigstrut}
\usepackage{array}
\usepackage{enumerate}
\usepackage{enumitem}
\usepackage[small,bf]{caption}
\usepackage{subcaption}
\usepackage{diagbox}
\usepackage{graphicx}
\usepackage{mathbbol}
\usepackage{appendix}
\usepackage{verbatim}
\usepackage[top=2.5cm,bottom=2.5cm,left=3cm,right=3cm]{geometry}
\usepackage{makecell}
\usepackage{listings}
\usepackage{mathtools}
\usepackage{dsfont}
\usepackage{braket}
\usepackage{longtable}
\usepackage{threeparttable}
\usepackage{floatpag}
\usepackage[normalem]{ulem}
\usepackage{multicol}
\usepackage{hyperref}
\usepackage[capitalise]{cleveref} 
\newcommand{\rep}[1]{\ensuremath{\boldsymbol{#1}}}

\begin{document}

\begin{center}
{\bf 
{\LARGE Reinforcement Learning techniques for the flavor problem in particle physics}
}\\[8mm]
A.~Giarnetti$^{\,a}$\footnote{E-mail: \href{mailto:alessio.giarnetti@roma1.infn.it}{\texttt{alessio.giarnetti@roma1.infn.it}}},
D.~Meloni$^{\,b}$\footnote{E-mail: \href{mailto:davide.meloni@uniroma3.it}{\texttt{davide.meloni@uniroma3.it}}},
 \\
 \vspace{5mm}
$^{a}$\,{\it \small INFN, Sezione di Roma, Rome, Italy} \\
$^{b}$\,{\it \small Dipartimento di Matematica e Fisica, Università di Roma Tre, Rome, Italy} \\
\end{center}

\vskip 7mm      
\begin{center}
\textbf{Abstract}
\end{center}
\vspace{-.5em} 

This short review discusses recent applications of Reinforcement Learning (RL) techniques to the flavor problem in particle physics. Traditional approaches to fermion masses and mixing often rely on extensions of the Standard Model based on horizontal symmetries, but the vast landscape of possible models makes systematic exploration infeasible. Recent works have shown that RL can efficiently navigate this landscape by constructing models that reproduce observed quark and lepton observables. These approaches demonstrate that RL not only rediscovers models already proposed in the literature but also uncovers new, phenomenologically acceptable solutions.


\section{Introduction}

The flavor problem in particle physics has been a central focus of study since the inception of the Standard Model (SM). Many attempts to explain fermion masses and mixing involve SM extensions based on horizontal symmetries, who are responsible of making the entries of the mass matrices related in a non trivial way instead of being free parameters. A common feature of all these models is the assignment of particles to irreducible representation of the flavor group (the case of non-abelian symmetries), the assignment of a $U(1)$ charge to particles (models \' a la Froggatt-Nielsen \cite{Froggatt:1978nt})
or a combination of both. The space of all possibilities is usually immense, spanning $\sim 10^n$ different models, with $n$ a large integer of $\mathcal{O}(10)$.
It is clear that the success of an SM extension strongly relies on the ability to select a realization that fits the low energy data; while human searches have led to a good success, it is not sure whether the discovered solutions are the minimal ones in terms of new degrees of freedom or the best ones according to other selection criteria like, for instance, the absence of huge hierarchies on the model parameters or a strong fine-tuning. The vastness of the parameters space in many fields of particle physics is one of the reason why, in very recent years, Machine Learning (ML) techniques started to be applied to several problems in Physics.
Entering into the details of all attempts performed using ML techniques in Physics is a formidable task, as in many areas a long list of papers exist and revising them is well beyond the scope of this short review.
In the restricted domain of theoretical physics, the subject is currently very active, covering a wide range of topics
 ranging (among others) from string theory \cite{Halverson:2019tkf,Mutter:2018sra,He:2017aed,Krefl:2017yox, Ruehle:2017mzq,Abel:2014xta,Matchev:2024ash,Larfors:2020ugo, Krippendorf:2021uxu, Constantin:2021for, Cole:2021nnt, Seong:2024wkt, Harvey:2024lwl, Lanza:2024mqp, Choi:2023rqg, Hirst:2022jfq, Cui:2022cxe, Bao:2022rup, Berman:2021mcw, Gao:2021xbs, Carifio:2017nyb, Liu:2017dzi, Wang:2018rkk, Altman:2018zlc, Jinno:2018dek, Bull:2018uow, Jejjala:2019kio, Brodie:2019dfx, Suresh:2024dsg,Bies:2020gvf, Ruehle:2020jrk,Anderson:2020hux, Cipriani:2025ini} to  QCD \cite{Shanahan:2018vcv,BarreraCabodevila:2025iwr,BarreraCabodevila:2025ogv,Favoni:2020reg,Favoni:2025ghc,Holland:2025fsa,Zhu:2025pmw,Aarts:2025gyp,Tomiya:2025quf,Christ:2024ydh,Gao:2024nzg,Apte:2024vwn,Cruz:2024grk,Abbott:2024kfc,Chatterjee:2024pbp,Holland:2024muu,Gao:2023quv,Gao:2023uel,Ermann:2023unw,Zhou:2023pti,He:2023zin,Lehner:2023bba,Shanahan:2022ifi,Urban:2022vjz,Boyda:2022nmh,Gerasimeniuk:2021cql}, not to mention statistical mechanics \cite{Carleo:2016svm,Wetzel:2017ooo,Schlomer:2025txn,Du:2025qph,Wetzel:2025uhj,Zhang:2024mqd,Suresh:2024nek,Cybinski:2024wgx,Sadoune:2024hxl,Kairon:2023adn,Jianmin:2023iyz,Baul:2023qmu,Naravane:2023hbr,Arnold:2023tsr,Ghosh:2023hsg,Schlomer:2023bbd,Chung:2023uuj,Johnston:2022cgm,Patel:2022clf,Sancho-Lorente:2021yko,Chung:2021eoy,Kim:2021ulz,Cole:2020hjx,Dawid:2020itb,Alexandrou:2019hgt,Dong:2019ypp,Giannetti:2018vif,Nomura:2017tgw}, thermodynamics \cite{Torlai:2016szv,ShibaFunai:2018aaw,Seif:2019wdb,Zhang:2024wgr,Gu:2022gcn}, the foundation of quantum mechanics \cite{Deng:2017uik,Chen:2022ytr,Bharti:2020gua,Gonzalez:2020frf,Alves:2024wwu,Deng:2017hws,Bharti:2019xdc,Canabarro:2018xwk,Krivachy:2020ggy} and cosmology \cite{Rudelius:2018yqi,Jejjala:2020wcc,Saxena:2024rhu,Ocampo:2024fvx,Savchenko:2024zyt,Decant:2024bpg,Savchenko:2025jzs,Hai-LongHuang:2025vfs,Eckner:2025waa,Villaescusa-Navarro:2021pkb,Hassan:2021ymv,Modi:2021acq,Modi:2020dyb,Dai:2020ekz,Taylor:2019mgj,Schmittfull:2017uhh,Flauger:2022hie}. 
In this short review, we focus more strictly on ML in particle physics \cite{Kawai:2024pws,Schwartz:2021ftp,Abdelhaq:2025nch,Bakshi:2025fgx,Boto:2025ovp}, with the intent of exploring Beyond the Standard Model theories \cite{Wojcik:2024lfy,Wojcik:2024azu,Chekanov:2025pmb} and, in particular,  the (a few, so far) efforts that have been done to implement AI agents in the study of the flavor problem. 
The subject is relatively young so, at least in principle, there exists a large probability that more can be said and done in the near future. This review would help in setting the state of the art on the attempts accomplished using the Reinforcement Learning (RL) technique \cite{Sutton} to find promising quark  \cite{Harvey:2021oue} and lepton \cite{Nishimura:2020nre,Baretz:2025zsv} textures\footnote{A study based on the Diffusion Model (DF) has been carried out for leptons \cite{Nishimura:2025knz}.}
addressing, at the same time, the quest for an axion candidate. More generally, the same RL-based strategy can be straightforwardly adapted to other areas of particle physics, such as to problems in theoretical physics where model building amounts to exploring large, discrete or continuous spaces under theoretical and phenomenological constraints. Among others, we mention string theory \cite{Abel:2021rrj}, particle jets physics \cite{Carrazza:2019efs}, lattice QCD \cite{Alvestad:2023jgl} and quantum field theory \cite{Windisch:2019byg, Zeng:2025xbh}.

The main features of RL are described in Sect.(\ref{rlintro}), and its interesting results for model building are analysed in sect.(\ref{rlanalisi}). Sect.(\ref{conc}) is devoted to our conclusions.

\section{Reinforcement Learning}
\label{rlintro}
Reinforcement Learning \cite{Sutton} is a subfield of Machine Learning that focuses on how an \emph{agent} learns to make optimal decisions by interacting with an \emph{environment} to maximize a cumulative \emph{reward} over time. Unlike supervised learning, where a model is trained on a dataset of labeled input-output pairs, or unsupervised learning, which seeks patterns in unlabeled data, RL operates in a trial-and-error framework, like a human strategist attempting to master a complex task. The agent learns by taking actions, observing their consequences, and receiving feedback in the form of rewards or penalties. This feedback loop allows the agent to gradually refine its decision-making strategy without being explicitly told what the ``correct'' action is.
This continuous refinement process, based on performance feedback rather than explicit instruction, mirrors how humans develop expertise in complex domains. This paradigm is well-suited for problems where the optimal course of action is not immediately clear, and the agent must explore its environment to discover effective strategies. RL is often formalized using \emph{Markov Decision Processes (MDPs)}, which provide a mathematical framework to model sequential decision-making under uncertainty. In RL, the agent observes the current \emph{state} of the environment, selects an \emph{action}, and receives a \emph{reward} along with the next state, iteratively learning a \emph{policy}---a strategy that maps states to actions---to maximize the expected sum of future rewards.

\subsection{Common features}

Although many variants of RL exixt, we can identify some common features; these components are essential for understanding how RL systems operate:

\begin{itemize}
    \item \textbf{Agent}: the entity responsible for making decisions and learning from experience. The agent could be a robot, a software program, or any system capable of taking actions in an environment;
    \item \textbf{Environment}: the external system with which the agent interacts. The environment includes everything outside the agent, such as a physical space, a game board, or a simulated market. It responds to the agent's actions by providing new states and rewards;
    \item \textbf{State} ($s \in \mathcal{S}$): a representation of the current situation or configuration of the environment at a given time. The state contains all relevant information the agent needs to make a decision. For example, in a chess game, the state could represent the positions of all pieces on the board. With $\mathcal{S}$ we indicate the ensemble of the whole states, whose single realization is indicated with $s$;
    \item \textbf{Action} ($a \in \mathcal{A}$): a decision or move the agent can make to influence the environment. Actions can be discrete (e.g., moving left or right) or continuous (e.g., adjusting the speed of a vehicle). The set of all possible actions $\mathcal{A}$ depends on the environment and the current state;
    \item \textbf{Reward} ($r \in \mathbb{R}$): a scalar value provided by the environment after each action, indicating the immediate benefit or cost of that action. The reward serves as feedback, guiding the agent toward desirable behaviors. For instance, a positive reward might be given for winning a game, while a negative reward (penalty) could be assigned for making an invalid move in a chess game;
    \item \textbf{Policy} ($\pi$): the strategy or rule that the agent follows to select actions based on the current state. Formally, a policy is a mapping from states to actions, denoted as $\pi: \mathcal{S} \rightarrow \mathcal{A}$ (for deterministic policies) or $\pi(a|s)$ (for probabilistic policies, which assign probabilities to actions given a state). The goal of RL is to learn an optimal policy that maximizes long-term rewards;
    \item \textbf{Value Function} ($V^\pi(s)$): a function that estimates the expected cumulative reward (also called the \emph{return}) an agent can achieve starting from a given state $s$ and following the policy $\pi$. The value function quantifies the long-term desirability of a state under a specific policy. A related concept is the \emph{action-value function} ($Q^\pi(s, a)$), which estimates the expected return for taking action $a$ in state $s$ and following policy $\pi$ afterward. Notice that the reward $r$ and the value function are fundamentally related: the reward provides the immediate feedback signal, while the value function is the estimation of the long-term benefit derived from those rewards.
    \item \textbf{Discount Factor} ($\gamma \in [0, 1]$): a parameter that determines the importance of future rewards relative to immediate ones. A value of $\gamma$ close to 0 prioritizes short-term rewards, while a value close to 1 emphasizes long-term rewards, promoting long-term planning by the agent.
    \item \textbf{Learning Rate} ($\alpha \in [0,1]$): a parameter which controls how much newly acquired information overrides old information when updating estimates. When $\alpha$ is close to 1, the agent quickly adapts to new experiences but tends to forget past knowledge; conversely, for $\alpha$ close to 0 the agent learns slowly, giving more weight to prior knowledge and less to new data. It can decay over time so to make updates smaller as learning converges.
\end{itemize}

As mentioned above, RL problems are typically formalized using MDPs, which provide a structured way to model sequential decision-making in environments with stochastic dynamics. An MDP is defined by a tuple $(\mathcal{S}, \mathcal{A}, P, \mathbb{R}, \gamma)$, where $(\mathcal{S}, \mathcal{A},\mathbb{R})$ and $\gamma$ have been defined above, and $P(s'|s, a)$ is the \emph{transition probability function}, which specifies the probability of transitioning from the current state $s$ to a new state $s'$ after taking action $a$.

The objective of RL is to find an \emph{optimal policy} $\pi^*$ that maximizes the \emph{expected discounted return}, defined as the expected sum of discounted rewards over an infinite horizon \cite{Sutton}:

\begin{equation}
G_t = \left[ \sum_{k=0}^{\infty} \gamma^k r_{t+k+1} \right]\,.
\label{gidit}
\end{equation}

Here, $G_t$ represents the \emph{return} starting from time step $t$, which is the total discounted reward accumulated by following policy $\pi$. 
To achieve this goal, RL algorithms often rely on two key approaches:
\begin{enumerate}
    \item \textbf{Value-based methods}: these methods estimate the value function ($V^\pi(s)$ or $Q^\pi(s, a)$) and derive a policy by selecting actions that maximize the estimated value. A classic example is \emph{Q-learning}, which iteratively updates the action-value function to learn an optimal policy without explicitly modeling the environment's transition dynamics \cite{Wang:2021yix}. At the heart of \emph{Q}-learning is the Bellman equation used to update \emph{Q}-values:

\begin{equation}
Q(s_t, a_t) \leftarrow Q(s_t, a_t) + \alpha \left[ r_{t+1} + \gamma \max_{a} Q(s_{t+1}, a) - Q(s_t, a_t) \right]\,,
\label{one}
\end{equation}
where $s_t$ is the current state at time $t$, $a_t$ is the action taken at time $t$,
$r_{t+1}$ is the reward received after taking action $a_t$, $s_{t+1}$ is the new state after taking action $a_t$, $\alpha$ is the learning rate,
 $\gamma$ is the discount factor, $\max_{a} Q(s_{t+1}, a)$ is the estimate of the maximum future reward from the next state.
In this procedure, one first initializes the \emph{Q}-table arbitrarily for all state-action pairs $Q(s, a)$; then, choose an action $a$ from state $s$ using a policy (e.g., $\varepsilon$-greedy\footnote{\label{fnlabel}\emph{Q}-learning balances exploration and exploitation typically using an $\varepsilon$-greedy policy:
\begin{itemize}
    \item With probability $\varepsilon$, choose a random action (exploration).
    \item With probability $1 - \varepsilon$, choose the action that maximizes the current \emph{Q}-value (exploitation).
\end{itemize}}), takes action $a$, observes reward $r$ and the next state $s'$. Then, the \emph{Q}-value is updated
            \begin{equation}
            Q(s, a) \leftarrow Q(s, a) + \alpha \left[ r + \gamma \max_{a'} Q(s', a') - Q(s, a) \right]
            \end{equation}
and one sets $s \leftarrow s'$.
The goal is to find the optimal \emph{Q}-function $Q^*(s, a)$ such that:
\begin{equation}
Q^*(s, a) = \mathbb{E} \left[ r_{t+1} + \gamma \max_{a'} Q^*(s_{t+1}, a') \,\middle|\, s_t = s, a_t = a \right]\,,
\end{equation}
where the symbol $\mathbb{E}$ defines the expected value.
In other words: the optimal value of taking action $a$ in state $s$ equals the expected value of the immediate reward $r_{t+1}$ plus the discounted ($\gamma$) value of the best possible future action in the next state $s_{t+1}$. The term "$\max_{a'}$" indicates that, in the future, the agent will always select the best action.

Once $Q^*(s, a)$ is learned, the optimal policy $\pi^*$ can be derived as:
\begin{equation}
\pi^*(s) = \arg\max_a Q^*(s, a) \,.
\end{equation} 
Notice that,  while $\max_a Q^*(s,a)$
gives the maximum value of the function $Q^*(s,a)$ over all possible actions $a$, 
with the notation $\arg\max_a Q^*(s,a)$ we denote the function which returns the action $a$ that achieves this maximum value.

\item \textbf{Policy-based methods}: 
These methods directly parameterize and optimize the policy $\pi_\theta(a|s)$, where $\theta$ are tunable parameters of the policy (e.g., weights of a neural network) \cite{Sequeira:2024htb}. Instead of estimating a value function first, policy-based methods adjust $\theta$ to maximize the expected return:
    \begin{equation}
    J(\theta) = \mathbb{E}_{\pi_\theta} \left[ \sum_{t=0}^\infty \gamma^t r_{t+1} \right].
\end{equation}
    The gradient of $J(\theta)$ with respect to the parameters can be estimated using sampled trajectories from the environment. A prototypical algorithm is \emph{REINFORCE}, which updates the parameters in the direction of the return:
    \begin{equation}
    \theta \leftarrow \theta + \alpha \sum_{t=0}^{\infty} \nabla_\theta \ln \pi_\theta(a_t|s_t) G_t,
    \end{equation}
    where $G_t$ is the cumulative discounted reward from time $t$ defined in eq.(\ref{gidit}). \emph{REINFORCE} is suitable for RL because it estimates the policy gradient from sampled trajectories without requiring a model of the environment. However, its gradient estimates have high variance, often necessitating variance-reduction techniques for stable and efficient learning.

    Policy-based methods naturally handle high-dimensional or continuous action spaces and can learn stochastic policies, which are useful in partially observable or highly uncertain environments.
\end{enumerate}
There also exist hybrid approaches called \textit{actor-critic} methods which combine value-based and policy-based ideas by learning both a value function (critic) and a policy (actor) to improve sample efficiency \cite{Kolle:2024lpx}.

In the following section, we will review the attempts to discuss the flavor problem by means of RL techniques. 

\section{Quark and lepton masses and mixing from RL}
\label{rlanalisi}


\subsection{$U(1)$ Froggatt-Nielsen for quark observables}
The first paper we want to discuss is Ref. \cite{Harvey:2021oue}, which explores the application of RL to model building, specifically focusing on Froggatt-Nielsen (FN) type models for quark masses and mixing. The primary goal is to investigate whether RL techniques can be used to train a neural network \cite{Nielsen} to construct particle physics models with certain prescribed properties. More specifically, the authors aimed to develop a system where an RL agent could efficiently lead from random, typically physically unacceptable FN models, to phenomenologically viable ones that are consistent with observed quark masses and mixing. In addition, they also wanted to test if the trained networks could not only discover new models but also identify models already proposed in the literature.

From the model building point of view, FN models are well known and, for the present paper, work as follows. In the Standard Model (SM), quark masses and mixings originate from Yukawa couplings:
\begin{equation}
    \mathcal{L}_{\rm Yuk} = Y^u_{ij} \overline{Q}^i H^c u^j + Y^d_{ij} \overline{Q}^i H d^j + \text{h.c.},
\end{equation}
where $Q^i$ are left-handed quark doublets, $u^i$ and $d^i$ are right-handed up- and down-type quarks, and $H$ is the Higgs doublet. After electroweak symmetry breaking, the Higgs acquires a VEV $\langle H^0 \rangle = v$, yielding the mass matrices:
\begin{equation}
    M_u = v Y^u, \qquad M_d = v Y^d.
\end{equation}
These are diagonalized via unitary matrices:
\begin{equation}
    M_u = U_u \hat{M}_u V_u^\dagger, \quad M_d = U_d \hat{M}_d V_d^\dagger,
\end{equation}
with $\hat{M}_u = \text{diag}(m_u, m_c, m_t)$ and $\hat{M}_d = \text{diag}(m_d, m_s, m_b)$. The observable mixing is encoded in the CKM matrix:
\begin{equation}
    V_{\rm CKM} = U_u^\dagger U_d,
\end{equation}
parameterized by three angles $\theta_{12}, \theta_{13}, \theta_{23}$ and a CP-violating phase $\delta$:
\begin{equation}
    V_{\rm CKM} = \begin{pmatrix}
        c_{12}c_{13} & s_{12}c_{13} & s_{13}e^{-i\delta} \\
        -s_{12}c_{23} - c_{12}s_{23}s_{13}e^{i\delta} & c_{12}c_{23} - s_{12}s_{23}s_{13}e^{i\delta} & s_{23}c_{13} \\
        s_{12}s_{23} - c_{12}c_{23}s_{13}e^{i\delta} & -c_{12}s_{23} - s_{12}c_{23}s_{13}e^{i\delta} & c_{23}c_{13}
    \end{pmatrix}.
\end{equation}

The experimentally measured parameters used in the paper are shown in Table~\ref{tab:exp} for reference.
\begin{table}[t!]
\centering
\begin{tabular}{|c|c|c|c|c|c|}
\hline
$m_u$ & $m_d$ & $m_c$ & $m_s$ & $m_t$ & $m_b$ \\
\hline
$0.00216^{+0.00049}_{-0.00026}$ & $0.00467^{+0.00048}_{-0.00017}$ & $1.27\pm0.02$ & $0.093^{+0.011}_{-0.005}$ & $172.4 \pm 0.07$ & $4.18^{+0.03}_{-0.02}$ \\
\hline
\end{tabular}

\vspace{2mm}

\begin{tabular}{|c|c|c|c|c|}
\hline
$v$ & $s_{12}$ & $s_{13}$ & $s_{23}$ & $\delta$ \\
\hline
$\sim 174$ & $0.22650 \pm 0.00048$ & $0.00361^{+0.00011}_{-0.00009}$ & $0.04053^{+0.00083}_{-0.00061}$ & $1.196^{+0.045}_{-0.043}$ \\
\hline
\end{tabular}
\caption{\it Reference values of the quark masses (in GeV) and CKM parameters used in \cite{Harvey:2021oue}.}
\label{tab:exp}
\end{table}
FN models introduce a flavor structure into the SM  via two main ingredients:
\begin{itemize}
    \item adding global symmetries $U_p(1)$, $p=1,\ldots,r$ to the SM gauge group;
    \item adding scalar singlets $\phi_p$, with $q_p(\phi) \neq 0$ responsible for the spontaneous breaking of $U_p(1)$'s.
\end{itemize}
The SM Yukawa terms are generally replaced by higher-dimensional operators:
\begin{equation}
    \mathcal{L}_{\rm Yuk} = \sum_{i,j} \left(
        a_{ij} \prod_p \phi_p^{n_{p,ij}} \overline{Q}^i H^c u^j +
        b_{ij} \prod_p \phi_p^{m_{p,ij}} \overline{Q}^i H d^j
    \right) + \text{h.c.},
\label{o1coeff}
\end{equation}
where $n_{p,ij}, m_{p,ij} \in \mathbb{Z}_{\geq 0}$; if the sum of all charges in a given operator is vanishing, then 
the related entries in the Yukawa matrix are different from zero. 
After the singlets acquire VEVs $v_p = \langle \phi_p \rangle$, the Yukawa matrices become:
\begin{equation}
    Y^u_{ij} = a_{ij} \prod_p v_p^{n_{p,ij}}, \qquad
    Y^d_{ij} = b_{ij} \prod_p v_p^{m_{p,ij}}\,,
\end{equation}
where $a_{ij}, b_{ij}$ are $\mathcal{O}(1)$ coefficients not determined by the flavor symmetries and fixed to some random number at the beginning of the simulation.
In this framework, a FN model is defined by vectors
\begin{equation}
\label{eqQ}
    \mathcal{Q}_p = \left(q_p(Q^i), q_p(u^i), q_p(d^i), q_p(\phi)\right),
\end{equation}
organized into an $r \times 10$ charge matrix $\mathcal{Q}$, where $r$ is the number of $U(1)$ symmetries, with bounded entries:
\begin{equation}
    q_{\rm min} \leq q_p \leq q_{\rm max}
\end{equation}
for each particle in the model.
The Higgs charges have been fixed to ensure the top Yukawa coupling is unsuppressed:
\begin{equation}
    q_p(H) = q_p(u^3) - q_p(Q^3).
\end{equation}
The FN landscape consists of many possible charge assignments. For example, with one $U(1)$ and $q_{\rm max} = - q_{min} = 9$, there are $19^{10}\sim 10^{13}$ models; for two $U(1)$ symmetries, $\sim 10^{26}$. Identifying viable models matching experimental data motivates the use of RL techniques.
The setup involves a single neural network $\pi_\theta$, parameterized by weights $\theta$, which defines the policy $\pi$. The network takes states as input and outputs a probability distribution over possible actions. Exploration of the environment is driven by this policy, meaning actions are sampled according to $\pi_\theta$ at each step in an {\it episode}, defined as the sequence of states $s_t$ and actions $a_t$ for $t=0,1,2,...$ (the maximum $t$ representing the length of the episode):
\begin{equation}\label{eppi}
 s_0\stackrel{\pi_\theta}{\xrightarrow{\hspace*{8mm}}} s_1\stackrel{\pi_\theta}{\xrightarrow{\hspace*{8mm}}}s_2\stackrel{\pi_\theta}{\xrightarrow{\hspace*{8mm}}}s_3\cdots \; .
\end{equation}

Data is collected by repeatedly running such episodes, implying that the system effectively consists of a single agent. According to the \textit{policy gradient theorem}, the policy network $\pi_\theta$ is trained to minimize the following loss function:

\begin{equation}\label{loss}
 L(\theta)=Q_\pi(s,a)\ln (\pi_\theta(s,a))\; .
\end{equation}

In practice, the action-value function $Q_\pi(s,a)$ is often approximated using the return $G$ observed from state $s$. The overall algorithm proceeds as follows:

\begin{enumerate}
    \item[(1)] Initialize the policy network $\pi_\theta$.
    \item[(2)] Collect a batch of data tuples $(s_t, a_t, G_t)$ by running multiple episodes as described in~\eqref{eppi}, with each episode starting from a randomly selected initial state $s_0$.
    \item[(3)] Update the policy network parameters $\theta$ using the collected data and the loss function in~\eqref{loss}.
    \item[(4)] Repeat steps (2) and (3) until convergence, i.e., when the loss becomes sufficiently small and the policy stabilizes.
\end{enumerate}
The authors model the FN theory space as a MDP. The state space $\mathcal{S}$ consists of all FN charge matrices $\mathcal{Q}$ of eq.(\ref{eqQ}). The action space $\mathcal{A}$ is composed of elementary updates that increment or decrement a single charge by one, that is ${\mathcal{Q}_p} \stackrel{a}{\longrightarrow} {\mathcal{Q}_p} \pm 1$. Since these updates are deterministic, the transition probabilities are not needed. There are $20\,r$ such actions in total. The discount factor is fixed to $\gamma = 0.98$.

In order to define a reward function, the authors first introduce the intrinsic value of a state $\mathcal{Q}$, defined as:
\begin{equation}\label{VQ}
\mathcal{V}(\mathcal{Q}) = -\min_{|v_p|\in I} \sum_\mu \left| \log_{10}\left(\frac{|\mu_{\mathcal{Q}, v_p}|}{|\mu_{\text{exp}}|} \right) \right| \; ,
\end{equation}
where the minimization is over scalar VEVs $v_p \in I\equiv [v_{\min}, v_{\max}]$ (with typical values $v_{\min} = 0.01$ and $v_{\max} = 0.3$) and $\mu$ runs over the six quark masses and CKM elements. This quantity measures the total deviation of the model predictions from the experimental values. In this context, a terminal state (that is a state after which the episode ends and a new one starts) satisfies $\mathcal{V}(\mathcal{Q}) > \mathcal{V}_0$ and all individual deviations (defined as $-\log_{10} |\mu_{\mathcal{Q}, v_p}|/|\mu_{\text{exp}}|$ computed for the $v_p$ value for which $\mathcal{V}(\mathcal{Q})$ is minimum) are greater than $ \mathcal{V}_1$, with $\mathcal{V}_0 = -10$ and $\mathcal{V}_1 = -1$.
Thus, a reward function is defined by:
\begin{equation}\label{reward}
\mathcal{R}(\mathcal{Q}, \alpha) = 
\begin{cases}
\mathcal{V}(\mathcal{Q}') - \mathcal{V}(\mathcal{Q}) & \text{if improvement,} \\
\mathcal{R}_{\text{offset}} & \text{otherwise}
\end{cases}
\end{equation}
with $\mathcal{R}_{\text{offset}} = -10$.
The training is performed using the \emph{REINFORCE} algorithm (described in the previous section) with ADAM optimization\footnote{ADAM is an optimization method that automatically adjusts how big each training step should be by keeping track of both the average direction of past steps (momentum) and how much those steps vary (adaptive learning rate), making it faster and more stable for training models.}. 
A single agent explores the FN environment using the policy $\pi_\theta$, running episodes of maximum length $N_{\text{ep}} = 32$ starting from random initial states. Terminal states discovered during training are stored for further analysis.
For models with one $U(1)$ symmetry, the trained network leads from random initial configurations to phenomenologically acceptable models in 93\% of episodes, with an average episode length 16.4. On the other hand, for models with two $U(1)$ symmetries, the network achieved a success rate of 95\% of episodes ending in terminal states. The average episode length in this more complex scenario was $\sim 20$. In particular, 4,630 unique terminal states were found for the one $U(1)$ symmetry case. Among these, 89 models had a high ``intrinsic value'' (meaning they fit experimental data well), with the best achieving a value of approximately $-0.598$. 57,807 unique terminal states were discovered for the two $U(1)$ symmetries case. Of these, 2,019 models showed high intrinsic value, with the best model having a value of approximately $-0.390$. Models with the highest intrinsic value for one and two $U(1)$ symmetries have been reported in Tab.\ref{intrvalue}. 
The training process, even for the larger two $U(1)$ symmetry environment, was achieved on a single CPU within reasonable timeframes of 1 hour for one $U(1)$ symmetry, and 25 hours for two $U(1)$ symmetries,  demonstrating the practical applicability of RL for navigating large model spaces where systematic scanning is impossible.
It is worth to mention that the trained networks were shown to be capable of guiding the search towards FN  models previously proposed in the literature, provided the search started from a nearby configuration in the model space.
\begin{longtable}{|c|c|}
\hline
charges & 
\resizebox{0.8\textwidth}{!}{$
\mathcal{Q}=\left(
\begin{array}{c|ccc|ccc|ccc|c|c}
{} & Q_1 & Q_2 & Q_3 & u_1 & u_2 & u_3 & d_1 & d_2 & d_3 & H & \phi \\ \hline 
q & 6 & 4 & 3 & -2 & 2 & 4 & -3 & -1 & -1 & 1 & 1
\end{array}
\right)
$} \\
\hline
$\mathcal{O}(1)$ coeff. & 
\resizebox{0.8\textwidth}{!}{$
(a_{ij}) \simeq
\begin{pmatrix}
 -1.975 & 1.284 & -1.219 \\
 1.875 & -1.802 & -0.639 \\
 0.592 & 1.772 & 0.982
\end{pmatrix}
\quad
(b_{ij}) \simeq
\begin{pmatrix}
 -1.349 & 1.042 & 1.200 \\
 1.632 & 0.830 & -1.758 \\
 -1.259 & -1.085 & 1.949
\end{pmatrix}
$} \\
\hline
VEV, value & $v_1\simeq 0.224\;,\quad \mathcal{V}(\mathcal{Q})\simeq -0.598$ \\
\hline
\hline
charges & 
\resizebox{0.8\textwidth}{!}{$
\mathcal{Q}=\left(
\begin{array}{rrr|rrr|rrr|r|rr}
Q_1 & Q_2 & Q_3 & u_1 & u_2 & u_3 & d_1 & d_2 & d_3 & H & \phi_1 & \phi_2 \\ \hline
2 & 2 & 1 & -2 & 0 & 1 & -1 & 0 & 1 & 0 & 1 & 0 \\
1 & 0 & 0 & 0 & 0 & 0 & -1 & -1 & -2 & 0 & 0 & 1
\end{array}
\right)
$} \\
\hline
$\mathcal{O}(1)$ coeff. & 
\resizebox{0.8\textwidth}{!}{$
(a_{ij})\simeq
\begin{pmatrix}
 -1.898 & 0.834 & -0.587 \\
 -0.575 & -0.592 & 1.324 \\
 -1.123 & -1.265 & 0.982
\end{pmatrix}
\quad
(b_{ij})\simeq
\begin{pmatrix}
 -1.759 & 1.358 & 1.013 \\
 -1.267 & 1.897 & -1.196 \\
 1.771 & 1.386 & -1.785
\end{pmatrix}
$} \\
\hline
VEVs, value & $(v_1,v_2)\simeq (0.079,0.112)\;,\quad\mathcal{V}(\mathcal{Q})\simeq -0.390$ \\
\hline
\caption{\it Examples of models with highest intrinsic value for one (top) and two (bottom) $U(1)$ symmetries. Charges and $\mathcal{O}(1)$ coefficients of eq.(\ref{o1coeff}) are given. Adapted from \cite{Harvey:2021oue}.\label{intrvalue}}\\
\end{longtable}

Even though not strictly related to FN models, we want to mention here the efforts provided in \cite{Matchev:2024ash} where machine learning techniques have been applied to construct particle physics models again in the Yukawa quark sector, aiming to satisfy two complementary criteria: \emph{truth} and \emph{beauty}. 
Truth is defined as the ability of a model to reproduce all experimental observables sensitive to its parameters. In the quark sector, this entails fitting quark masses and mixing parameters, including the CKM matrix elements and the Jarlskog invariant. 
Beauty\footnote{The implementation of the beauty definitions in the context of machine learning approaches to BSM models in flavor physics might be useful to narrow down the viable models in order to search for the most theoretically appealing solutions.}, in the authors’ definition, refers to desirable theoretical attributes beyond mere data fitting. To quantify beauty in Yukawa matrices, three measures are tested, which are 
uniformity (that is all matrix elements have approximately equal magnitude), sparsity (where a significant number of elements vanish exactly or are negligibly small) and symmetry (in which off-diagonal elements are mirrored in magnitude across the main diagonal). Using pseudo-experiments, the optimization procedure efficiently navigates the huge parameter space, uncovering models that achieve both experimental accuracy and aesthetic elegance. This effort demonstrates how machine learning can be applied in particle physics, not only as an efficient method for exploring vast parameter spaces but also as a flexible tool for model selection (taking into account  theoretical and phenomenological constraints), thus bridging numerical optimization and model-building intuition.

\subsection{Extention of the $U(1)$ Froggatt-Nielsen  to lepton observables}

We now discuss the paper \cite{Nishimura:2020nre} which extends somehow the previous work   to investigate the flavor structure not only for quarks but also for leptons  using RL. Specifically, the authors apply a value-based RL algorithm to models with a $U(1)$ flavor symmetry based on the Froggatt-Nielsen (FN) mechanism.
While, as said, for $U(1)$ charges $q_i$ in the range $-9 \leq q_i \leq 9$ to quarks alone yields approximately $10^{13}$ possible configurations, the inclusion of the lepton sector expands the parameter space  to over $10^{24}$ possibilities. This combinatorial explosion renders brute-force searches infeasible and motivates the application of machine learning techniques for efficient model exploration.
The system is provided with the following components: 
\begin{itemize}
    \item \textbf{Deep \emph{Q}-Network (DQN):} a DQN implementation has been supplied to approximate the \emph{Q}-value function, guiding the agent's actions via a neural network. In fact, instead of storing a \emph{Q}-table, the DQN uses a neural network to approximate the \emph{Q}-function. It leverages experience replay (learning from past transitions randomly sampled from memory) to break correlations between consecutive experiences, and incorporates a target network to stabilize training by maintaining a delayed copy of the \emph{Q}-network for computing target values.
    \item \textbf{Network Architecture:} 
    The architecture of each network is composed of several fully connected (dense) layers, 
where each neuron in a layer is connected to every neuron in the subsequent layer. 
The activation function used in these layers is the Scaled Exponential Linear Unit (SELU), 
which helps maintain normalized activations and supports stable training. 
The final layer of the network uses a \textit{softmax} activation (that is, a function that converts raw outputs into probabilities summing to one) to produce a probability distribution 
over the possible actions.
\end{itemize}

As before, the training procedure is implemented, first, by inputting the current state $s$ (a charge vector $\mathcal{Q}_p$)  into the target network; then, actions $a$ are chosen via an $\epsilon$-greedy policy (see footnote \ref{fnlabel}).
After that, the transition $e = \langle s, a, s', {\cal R} \rangle$ (where ${\cal R}$ is the reward received by the agent, more on this later) is stored in a replay buffer.  A mini-batch of experiences is sampled to update the \emph{Q}-network using a Huber loss function, optimized via stochastic gradient descent. 
The Huber loss function is an  error measure that combines the 
advantages of the mean squared error (MSE) and the mean absolute error (MAE). 
For small differences between the predicted value $y_{\text{pred}}$ and the target value $y$, 
it behaves like MSE, while for large differences it behaves like MAE, 
increasing its numerical stability while still being sensitive to small errors.
Denoting the outputs of the \emph{Q}-network and the target network by $y_i$ and $y_i^\prime$, respectively, the weights and the biases are updated by minimizing:
\begin{align}
    \begin{split}
        L_{\rm Huber}(y, y') = 
        \left\{
        \begin{array}{l}
             \frac{1}{2} (y'_{\left( {\cal R} \right)i} - y_i)^2\qquad \qquad\,\,\,\, \text{if}\ |y'_{\left( {\cal R} \right)i} - y_i| \leq \delta   \\
             \delta \cdot |y'_{\left( {\cal R} \right)i} - y_i| - \frac{1}{2}\delta^2\qquad \text{if}\ |y'_{\left( {\cal R} \right)i} - y_i| > \delta
        \end{array}
        \right.
        ,
    \end{split}
    \label{eq:Huber}
    \end{align}
    with $y'_{\left( {\cal R} \right)} = {\cal R} + \gamma y',\ \gamma = 0.99$ and $\delta = 1$, which combines a mean squared error and a mean absolute error.

At the end, the \emph{Q}-network parameters $\Theta$ are softly copied to the target network parameters $\Theta'$ using a learning rate $\alpha$.
The models investigated by the authors are built with a  single complex scalar flavon field $\phi$ with a $U(1)$ charge. As usual, 
Yukawa couplings are suppressed by powers of $\langle \phi \rangle / \Lambda$, generating fermion mass hierarchies. In addition, the Higgs $U(1)$ charge is fixed to allow the top Yukawa term without flavon suppression.
The resulting fermion masses and mixings arise after $\phi$ and the Higgs acquire VEVs: $v_\phi$ and $v_{EW}$.
The total charge vector $\mathcal{Q}_p$ contains 19 components (for all fermions); each component lies within $[-9, 9]$, leading to $19^{19}\sim10^{24}$ possible charge configurations, as previously anticipated. The flavon charge $q(\phi)$ is set to $+1$ or $-1$ with equal probability. Agent actions consist of incrementing or decrementing a component of $\mathcal{Q}_p$ by one unit.

The relevant Yukawa terms for leptons (those for quarks are analogous to the previous section) are given by:
\begin{align}
    {\cal L} &=  y^l_{ij}\left(\frac{\phi}{M}\right)^{n^l_{ij}}\Bar{L}_i H l_j
              \nonumber\\
              &+ y^\nu_{ij}\left(\frac{\phi}{M}\right)^{n^\nu_{ij}}\Bar{L}_i H^c N_j
              + \frac{y^N_{ij}}{2}\left(\frac{\phi}{M}\right)^{n^N_{ij}}M\Bar{N}^{c}_i N_j
              + \text{h.c.},
\label{eq:Lagrangian}
\end{align}
where $\{L_i, l_i, N_i, H\}$ denote the left-handed leptons, 
the right-handed charged leptons, the right-handed neutrinos, and the SM Higgs doublet with $H^c = i\sigma_2 H^\ast$, respectively. 
In the paper,  the authors assume three right-handed neutrinos and that the physical neutrino masses are generated by Type-I seesaw mechanism where 
the large mass $M$ is chosen as $M=10^{15}$\,GeV and the Yukawa couplings (including those for u and d quarks) $\{y^l_{ij},y^\nu_{ij},y^N_{ij}\}$ are ${\cal O}(1)$ real coefficients.

The reward ${\cal R}$ is based on an \textit{intrinsic value function} ${\cal V(Q)}$, which quantifies how well a charge configuration reproduces the observed fermion masses and mixing angles:
\begin{align}
\label{intrinsic}
        {\cal V(Q)} =
        \begin{cases}
            \displaystyle - {\cal M}_{\rm quark}+ {\cal C}, &  \\[6pt]
            \displaystyle -\left[{\cal M}_{\rm lepton} +  {\cal M}_{\rm neutrino}+ {\cal P}\right]. & 
        \end{cases}
    \end{align}
In eq.(\ref{intrinsic}),  
${\cal M}_{\rm quark}$ and ${\cal M}_{\rm lepton}$  measure the deviation of the predicted quark and lepton masses from their corresponding experimental values,
    \begin{align}
        {\cal M}_{\rm quark} &= \sum_{\alpha \in \{u,d\}} E_\alpha, &
        {\cal M}_{\rm lepton} &= \sum_{\alpha \in \{l\}} E_\alpha,
    \end{align}
    where
    \begin{align}
        E_\alpha = \left| \log_{10} \left( \frac{|m_\alpha|}{|m_{\alpha,{\rm exp}}|} \right) \right|.
    \end{align}
In addition, since the ordering of neutrino masses is not yet experimentally established, they considered two approaches: RL with a fixed ordering (either Normal Ordering (NO) or Inverted Ordering (IO)) and RL without specifying the ordering. The intrinsic value relevant to the neutrino masses ${\cal M}_{\rm neutrino}$ is  given by:
    \begin{align}
        {\cal M}_{\rm neutrino} =
        \begin{cases}
            \displaystyle \sum_{\alpha \in \{21,31\}} E_\alpha^{\nu}, &  \\[6pt]
            \displaystyle \sum_{\alpha \in \{21,32\}} E_\alpha^{\nu}, & 
        \end{cases}
    \end{align}
    with
        $E_\alpha^{\nu} = \left| \log_{10} \left( \frac{|\Delta m_\alpha^2|}{|\Delta m_{\alpha,{\rm exp}}^2|} \right) \right|$. 
Finally, the intrinsic value also incorporates quark and lepton mixing information through the quantities 
        ${\cal C} = \sum_{i,j} E_{\cal C}^{ij},~
        {\cal P} = \sum_{i,j} E_{\cal P}^{ij},$
    where
    \begin{align}
        E^{ij}_{\mathcal{C}} &= \left| \log_{10} \left( \frac{|V_{\rm CKM}^{ij}|}{|V_{\rm CKM,\,exp}^{ij}|} \right) \right|, &
         E^{ij}_{\mathcal{P}} &= \left| \log_{10} \left( \frac{|V_{\rm PMNS}^{ij}|}{|V_{\rm PMNS,\,exp}^{ij}|} \right) \right|.
    \end{align}
  The experimental values of the lepton observables used by the authors are listed in Table~\ref{tab:data_lepton}.
\begin{table}[h!]
 \centering
\small
   \begin{tabular}{|c||c|c||c|c|} \hline
     \multirow{2}{*}{Observables} & \multicolumn{2}{c||}{Normal Ordering} & \multicolumn{2}{c|}{Inverted Ordering}  \\
     \cline{2-5}
       & $1\sigma$ range & $3\sigma$ range & $1\sigma$ range & $3\sigma$ range  \\
     \hline
     $\sin^{2}\theta_{12}$ & $0.303_{-0.011}^{+0.012}$ & $0.270\rightarrow 0.341$ & $0.303_{-0.011}^{+0.012}$ & $0.270\rightarrow 0.341$ \\
     \hline
     $\sin^{2}\theta_{13}$ & $0.02225_{-0.00059}^{+0.00056}$ & $0.02052\rightarrow 0.02398$ & $0.02223_{-0.00058}^{+0.00058}$ & $0.02048\rightarrow 0.02416$ \\
     \hline
      $\sin^{2}\theta_{23}$ & $0.451_{-0.016}^{+0.019}$ & $0.408\rightarrow 0.603$ & $0.569_{-0.021}^{+0.016}$ & $0.412\rightarrow 0.613$ \\
     \hline
      $\delta_{\text{CP}}/\pi$ & $1.29_{-0.14}^{+0.20}$ & $0.80\rightarrow 1.94$ & $1.53_{-0.16}^{+0.12}$ & $1.08\rightarrow 1.91$ \\
     \hline
     $\cfrac{\Delta m_{21}^{2}}{10^{-5} \mathrm{eV}^{2}}$ & $7.41_{-0.20}^{+0.21}$ & $6.82\rightarrow 8.03$ & $7.42_{-0.20}^{+0.21}$ & $6.82\rightarrow 8.04$ \\
     \hline
     $\dfrac{\Delta m_{3l}^{2}}{10^{-3} \mathrm{eV}^{2}}$ & $2.507_{-0.027}^{+0.026}$ & $2.427\rightarrow 2.590$ & $-2.486_{-0.028}^{+0.025}$ & $-2.570\rightarrow -2.406$ \\ \hline
     $m_{e}$/\text{MeV} & \multicolumn{4}{c|} {$0.510999$} \\ 
     \hline
     $m_{\mu}$/\text{MeV} & \multicolumn{4}{c|}{$105.658$} \\
     \hline
     $m_{\tau}/\text{MeV}$ & \multicolumn{4}{c|}{$1776.86$} \\
     \hline
 \end{tabular}
     \renewcommand{\arraystretch}{1}
\caption{\it Experimental values for the lepton sector obtained from global analysis of the data, 
where $\Delta m_{3l}^{2}\equiv \Delta m_{31}^{2}=m^2_3 -m^2_1>0$ for NO and $\Delta m_{3l}^{2}\equiv \Delta m_{32}^{2}=m^2_3 -m^2_2<0$ for IO. Taken from \cite{Nishimura:2020nre}.}
\label{tab:data_lepton}
\end{table}
A configuration is considered a \textit{terminal state} if:
\[
{\cal V(Q)} < V_0, \quad E_\alpha < V_1, \quad E_\nu, E_{ij}^{(C,P)} < V_2
\]
with typical thresholds $V_0 = 10.0$, $V_1 = 1.0$, and $V_2 = 0.2$. 

In the quark sector, approximately $10^{12}$ configurations were explored.
After $\sim$15 hours on a single CPU, the agent reached terminal states after 20,000 episodes.
More than 6\% of 100,000 episodes reached valid terminal states; 21 independent charge configurations consistent with experimental data were identified. Monte Carlo optimization of $\mathcal{O}(1)$ Yukawa coefficients refined the results to fit within $<0.1\%$ of observed values. However, as the authors claimed, no CP violation is present in this sector due to phase rotation freedom. 

\begin{figure}[h!]
    \centering
    \begin{subfigure}{0.35\textwidth}
        \includegraphics[width=\textwidth]{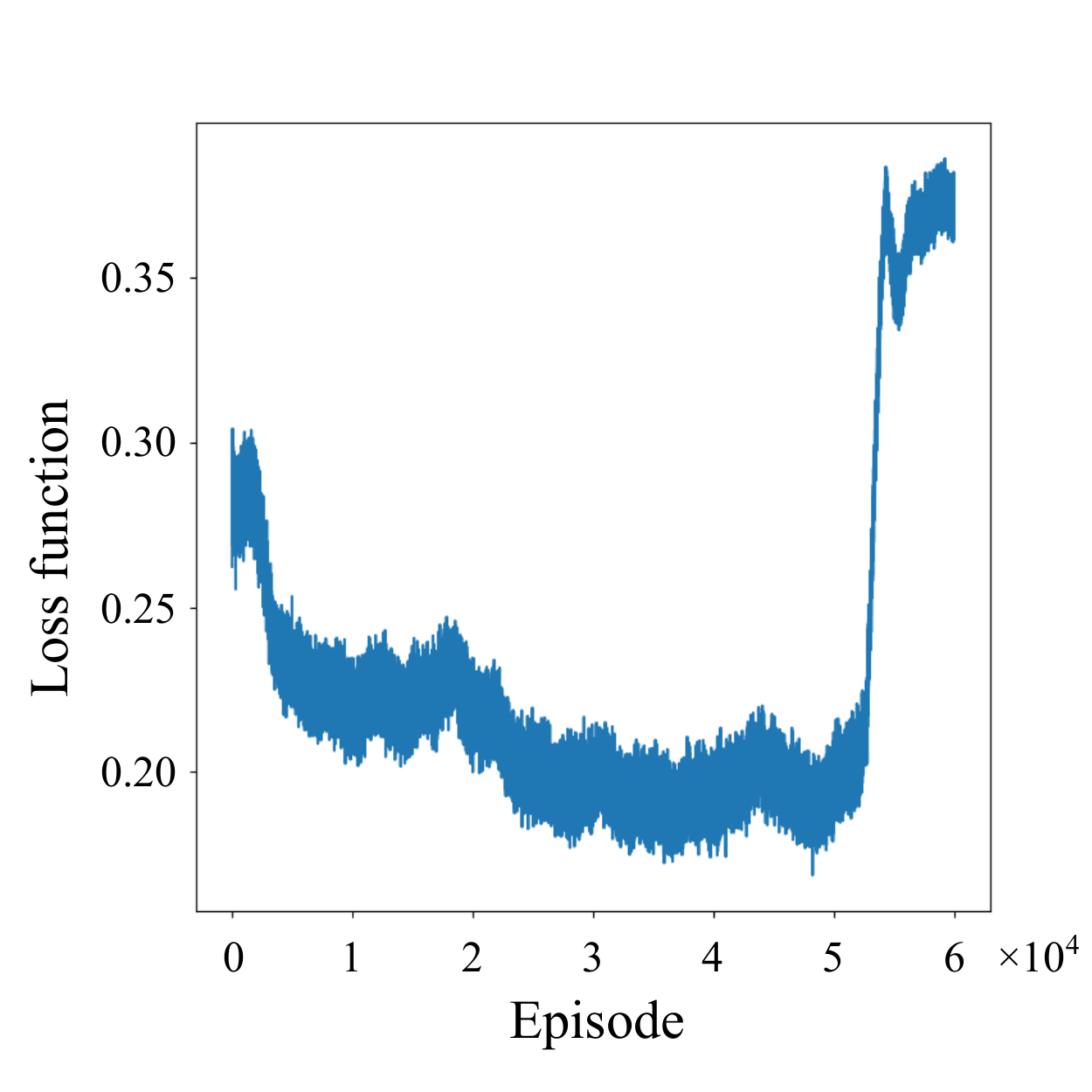} 
    \end{subfigure}\\
    \begin{subfigure}{0.35\textwidth}
        \includegraphics[width=\textwidth]{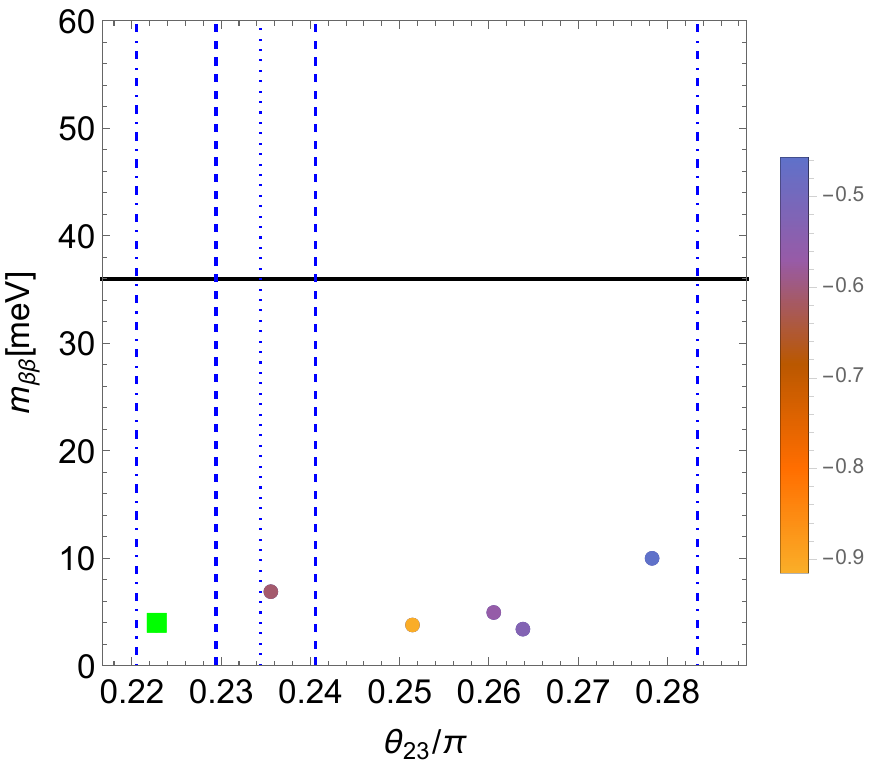}
    \end{subfigure}
    \begin{subfigure}{0.36\textwidth}
        \includegraphics[width=\textwidth]{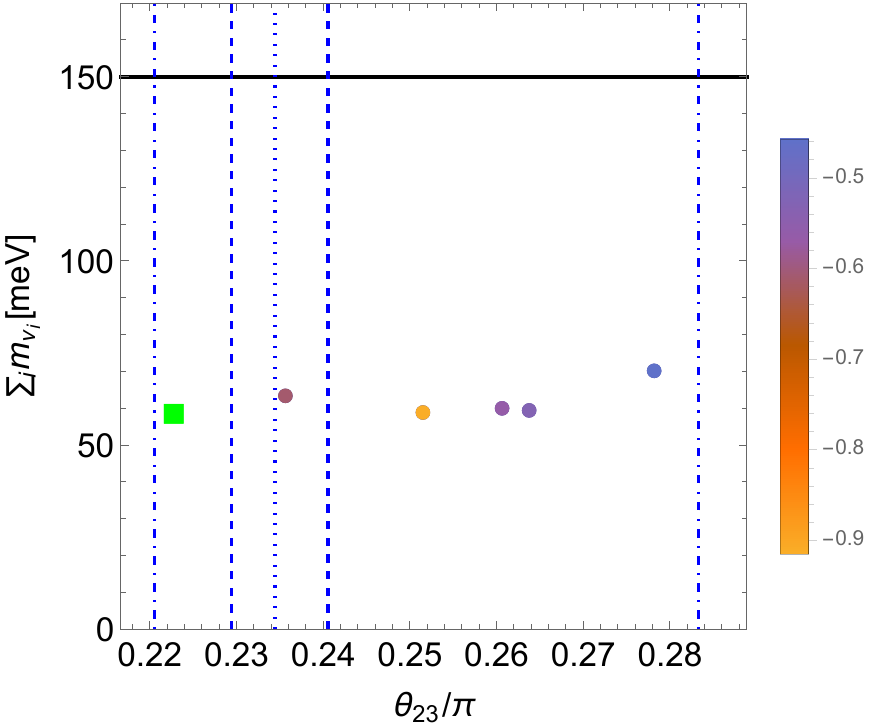}
    \end{subfigure}
    \caption{\it Learning results for the lepton sector by RL having specified the neutrino mass ordering. Upper panel: loss function vs episode number. Lower panels:  effective neutrino mass $m_{\beta\beta}$ (left plot) and sum of the neutrino masses (right plot) versus the atmospheric mixing angle $\theta_{23}$, for the 6 viable models found by the authors. 
    See text for further details.
    Adapted from \cite{Nishimura:2020nre}.}
\label{fig:result_lepton_unfixed}    
\end{figure}

In the lepton sector, the authors analyzed models in two different situations of mass ordering, one in which the ordering is fixed a priori and another when it is left free to be chosen by the agent action. For both cases, 
the Higgs charges and flavon VEVs were fixed from the quark sector. 
In the case of fixed mass ordering, the training started to find terminal states after roughly  5,000 episodes (around 8 hours on a single CPU) while the loss function tended to be minimized
around 50000 episodes, see upper panel of Fig.(\ref{fig:result_lepton_unfixed}). 
Here, 
the horizontal axis shows the number of training episodes, while the vertical axis reports the value of the loss function. A decreasing trend with fluctuations can be observed, indicating that the agent is gradually improving its predictions, with variations arising from the stochastic nature of the learning process. After $\sim 5\times 10^4$ episodes, the loss no longer decreases; instead, it slightly rises, suggesting potential overfitting to the specific experiences in the replay buffer.
In particular, 
terminal states appeared in 0.06\% of the cases: 63 (NO) and 121 (IO) total.  
Monte Carlo scans revealed  6 viable models for NO within 3$\sigma$ while no valid models for IO were found. These results suggest that NO is statistically preferred.
A typical successful output with a large intrinsic value is reported in Tab.\ref{tab:lepton_fixed_NO}, together with the lepton and scalar charges, the Yukawa matrices, the flavon vev, the obtained PMNS matrix and the predictions for the Majorana phases and $m_{\beta\beta}$.
\begin{table}[h!]
\centering
\resizebox{\textwidth}{!}{
\begin{tabular}{l|l}
\hline
Charges &
$\mathcal{Q}=
\left(\begin{array}{ccc|ccc|ccc|cc}
L_{1} & L_{2} & L_{3} & N_{1} & N_{2} & N_{3} & l_{1} & l_{2} & l_{3} & H & \phi \\ \hline
2 & 1 & 2 & -8 & -1 & -9 & -7 & -3 & -3 & -1 & 1
\end{array}\right)$ \\ \hline

$\mathcal{O}(1)$ coeff. &
$\begin{aligned}
y^{l} &\simeq
\begin{pmatrix}
 -0.889 & 2.056 & -0.299 \\
 -1.584 & -2.697 & 1.542 \\
 -0.797 & 0.918 & 1.501
\end{pmatrix}, \\
y^{\nu} &\simeq
\begin{pmatrix}
 1.135 & -1.331 & 0.128 \\
 1.207 & -1.203 & -0.051 \\
 -0.671 & -2.639 & 0.074
\end{pmatrix}, \\
y^{N} &\simeq
\begin{pmatrix}
 1.125 & -0.388 & 0.950 \\
 -0.388 & 1.066 & -0.349 \\
 0.950 & -0.349 & -0.656
\end{pmatrix}
\end{aligned}$ \\ \hline

VEV & $v_{\phi} \simeq 0.268\, e^{-0.166 i}$ \\ \hline
Intrinsic value & $\mathcal{V}_{\mathrm{opt}} \simeq -0.853$ \\ \hline

PMNS matrix &
$\left|V_{\mathrm{PMNS}}\right| \simeq
\begin{pmatrix}
0.819 & 0.553 & 0.154 \\
0.346 & 0.689 & 0.637 \\
0.458 & 0.468 & 0.756
\end{pmatrix}$ \\ \hline

Majorana phases & $\alpha_{21} \simeq 0.0,\quad \alpha_{31} \simeq -0.106\pi$ \\ \hline
Effective mass & $m_{\beta\beta} \simeq 3.793\ \mathrm{meV}$ \\ \hline
\end{tabular}
}
\caption{\it Highest intrinsic value scenario for the lepton sector with NO, where the neutrino mass ordering is specified in the network training. Adapted from \cite{Nishimura:2020nre}.}
\label{tab:lepton_fixed_NO}
\end{table}

For the sake of completeness, in the lower panels of Fig.\ref{fig:result_lepton_unfixed} we report 
the effective neutrino mass $m_{\beta\beta}$ (left plot) and the sum of the neutrino masses versus the atmospheric mixing angle $\theta_{23}$ (right plot) for the 6 viable models found by the authors. In the last two plots, the dotted line represents the global best fit value from NuFIT v5.2 results with Super-Kamiokande atmospheric data \cite{Esteban:2020cvm}, while the region inside the dashed (dot-dashed) line corresponds to $1\sigma$ ($3\sigma$) confidence level interval. 
The  square corresponds to the model listed in Tab.(\ref{tab:lepton_fixed_NO}).
The left panel shows that the predicted effective Majorana neutrino masses are all around a factor of 4 below its upper bound of
36 meV (90\% CL) indicated  by the black solid line
\cite{KamLAND-Zen:2022tow}; from the right panel, instead, we see that the sum of neutrino masses are all roughly 3 times smaller than the upper bound of $15$ meV (95\% CL), corresponding to the black solid line in the case of $\Lambda$CDM model \cite{RoyChoudhury:2019hls}.

In the case of unfixed mass ordering, 
the training  took $\sim$12 hours and 
terminal states where found in over 60\% of 60,000 episodes. In two dedicated runs, the authors found more that 13k terminal states for NO and  22k for IO. After optimization of the $\mathcal{O}$(1) coefficients, 15 NO models matched observations within 3$\sigma$ while, again, no IO models  were viable.
This work demonstrates that reinforcement learning can effectively explore the flavor structure of fermions in models with $U(1)$ symmetry. The results strongly suggest that Normal Ordering is statistically favored over Inverted Ordering in the neutrino sector.
\subsection{An AI agent for autonomous model building}
As a step forward in response to the challenges posed by the flavor problem in the $\nu$ sector, the authors of \cite{Baretz:2025zsv} introduce the \emph{Autonomous Model Builder} (AMBer), an AI-assisted framework based on RL. AMBer is designed to efficiently navigate the vast combinatorial landscape of model-building possibilities. In its implementation, the system autonomously selects symmetry groups, defines particle content, assigns group representations, and evaluates the resulting models. The goal is to construct viable models that both minimise the number of free parameters and  reproduce experimental observations. While the framework has been demonstrated in the context of neutrino flavor mixing, the underlying methodology is general and can be adapted to other domains of theoretical physics.

In the paper, the problem of constructing neutrino flavor models is reformulated as an RL task. The \emph{environment} in which the RL agent operates is defined by the space of possible models, with the current state corresponding to a specific model under consideration. Actions correspond to modifications of the model, such as altering particle content, changing group representations, or modifying symmetry properties, subject to mathematical consistency conditions. Each particle is encoded via a one-hot vector reflecting its non-Abelian representation and Abelian charge, while additional state information includes lepton triplet associations, vacuum expectation value configurations for flavons, and the order of the Abelian symmetry.

The agent evaluates its performance using a reward function carefully designed to encourage models that achieve a good fit to data, minimize the number of free parameters, and explore higher-order Abelian symmetries when advantageous. Invalid actions or models are penalized, while valid and predictive configurations are rewarded. The reward is normalized between $-1$ and $1$, with high-quality states boosted to maximize their influence on learning.

To enable rapid model evaluation, the framework incorporates an optimized software pipeline. A Python reimplementation of the Mathematica \texttt{Discrete} package—\texttt{PyDiscrete} computes Clebsch–Gordan coefficients with substantial speed gains. The \texttt{Model2Mass} package symbolically constructs the superpotential and extracts the charged-lepton, Dirac neutrino, and Majorana mass matrices for arbitrary finite non-Abelian groups. For parameter fitting, the \texttt{FlavorPy} package is employed, performing multiple short optimisation runs from varied initial conditions to avoid local minima while containing computational costs.

The RL algorithm chosen for AMBer is Proximal Policy Optimisation (PPO) that learns a policy directly and updates it carefully so it does not change too much at once. It uses a value function to help judge how good actions are, learns only from data generated by its current behavior, and uses advantage estimates to make learning more stable. By reusing the same data for a few updates, PPO achieves reliable learning without becoming unstable.
The loss function includes value, policy, and entropy terms, balancing exploitation and exploration. Training is performed in parallel over 250 environments, each with a maximum episode length of 1000 steps. Evaluation of models occurs every few steps, with the evaluation frequency depending on the chosen symmetry group. The search is conducted in several theory spaces, such as $A_4 \times Z_4$ with up to five flavons, $A_4 \times Z_N$ with variable $N$, and $T_{19} \times Z_4$ with up to six flavons.

\begin{figure}[t!]
    \centering
    \includegraphics[width=1.0\textwidth]{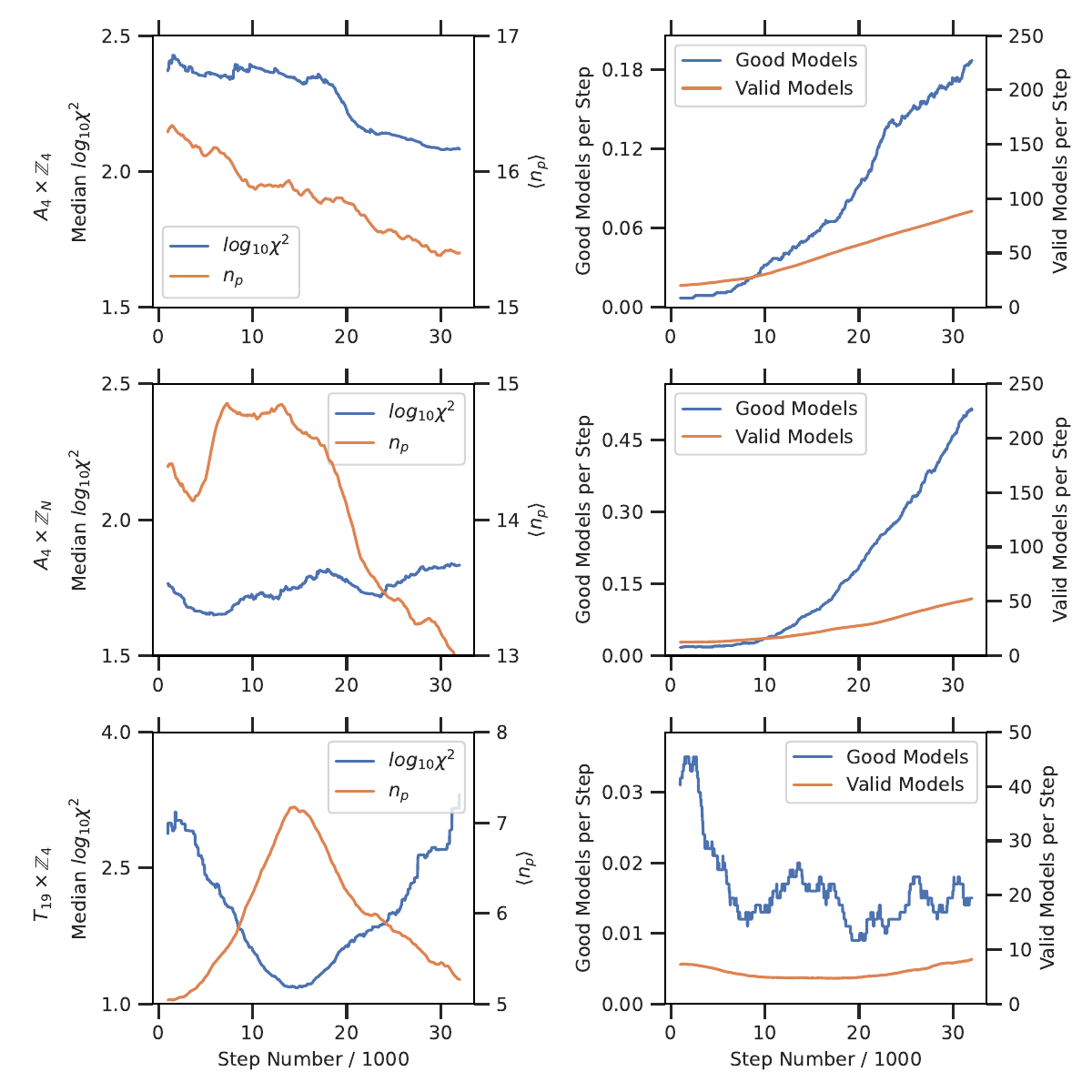}
    \caption{\it Training variables of interest over time for searches in three spaces: $A_4 \times  \mathbb{Z}_4$ (top), $A_4 \times \mathbb{Z}_N$ (middle), and $T_{19} \times \mathbb{Z}_4$ (bottom). Left column: evolution of $\chi^2$ in blue (where the curve indicates the median $\log_{10}{\chi^2}$ over all environments) and the mean number of parameters $\langle n_p \rangle$ as training progresses in orange. Right column: number of valid models in orange and good ($\chi^2 \leq 10$ and $n_p \leq$  7) models in blue. From \cite{Baretz:2025zsv}.}
    \label{fig:RewardDiagnostics}    
\end{figure}

Throughout training, the agent’s performance is monitored by tracking the $\chi^2$ fit to data, the number of free parameters, and other internal metrics, see Fig.(\ref{fig:RewardDiagnostics}).
Comparisons with random scans reveal that AMBer significantly outperforms naive exploration strategies, particularly in the $A_4$-based theory spaces.

The application of AMBer to neutrino flavor model building yields thousands of models that are both predictive (having at most seven free parameters) and in good agreement with experimental data, with $\chi^2 \leq 10$. In the $A_4 \times Z_4$ search space, many of the models rediscovered align closely with known constructions from the literature, featuring typical triplet assignments for lepton doublets and right-handed neutrinos, along with favoured vacuum alignments such as $(1,1,1)$. The $A_4 \times Z_N$ runs (see Tab.(\ref{tab:a4zn_found_model}) for a typical output) demonstrate the agent’s flexibility in selecting symmetry orders.
\begin{table}[t!]
\centering
\begin{tabular}{lllllllllll}
\toprule
  & $L$  & $E_1$ & $E_2$ & $E_3$& $N$ & $H_u$ & $H_d$ & $\phi_1$  & $\phi_2$ & $\phi_3$ \\ 
 \midrule
 $A_4$  & $\rep{3}$ & $\rep{1''}$ & $\rep{1'}$ & $\rep{1}$ & $\rep{3}$  & $\rep{1}$  & $\rep{1''}$ & $\rep{1'}$ & $\rep{3}$ & $\rep{3}$ \\ 
 $\mathbb{Z}_7$ & $1$ & $2$ & $2$ & $2$ & $2$ & $2$ & $2$ & $2$ & $5$ & $2$ \\
 \bottomrule
\end{tabular}
\caption{\it Moddel found by AMBer in the $A_{4}\times \mathbb{Z}_N$ search (with $N=7$), with 5 parameters and $\chi^2=0.79$. Here $\langle \phi_2 \rangle / \left( 0.1 \Lambda \right) = \langle\phi_3\rangle / \left( 0.1 \Lambda\right)
=  (1,1,1)$.}
\label{tab:a4zn_found_model}
\end{table}
In the larger and less charted $T_{19} \times Z_4$ theory space, AMBer uncovers models with more uniform representation assignments and an expanded variety of viable structures. While the relative gain over random scans is smaller here, AMBer still identifies a greater number of good models, including particularly simple configurations with very few free parameters.





\section{Conclusions}
\label{conc}
In this short review, we have summarized the first applications of Reinforcement Learning (RL) to the flavor problem in particle physics. Traditional approaches to quark and lepton mass hierarchies face the difficulty of exploring a vast model space, which quickly becomes intractable with increasing symmetries and parameters. RL-based frameworks have demonstrated the ability to efficiently navigate these spaces, identifying viable Froggatt--Nielsen charge assignments for quarks and leptons, often rediscovering known models and, in some cases, uncovering new ones consistent with experimental data. In the lepton sector, RL results point toward a statistical preference for Normal Ordering of neutrino masses. More advanced frameworks, such as AMBer, further illustrate the potential of AI agents to autonomously construct Beyond Standard Model theories with minimal parameter dependence and predictive power. Although this field is still in its infancy, the promising results obtained so far indicate that RL may become a valuable tool in theoretical model building, complementing traditional approaches and opening new directions in the search for solutions to the flavor problem,  as those based on modular symmetries and non-holomorphic modular forms.

In the  future, we think that RL will have a large impact on the development of fully autonomous theory-design frameworks and that its ability in exploring more complex parameter spaces against simple random scan  will be clarified.

\newpage

\bibliographystyle{JHEPwithnote}
\bibliography{bibliography}

\end{document}